\documentclass{ws-ijsc}

\pdfoutput=1
\def\draftnote{Preprint}

\usepackage[english]{babel}

\usepackage[utf8]{inputenc}

\usepackage[T1]{fontenc}
\usepackage{textcomp}

\usepackage[table]{xcolor}
\usepackage{graphicx}

\usepackage[]{csquotes}

\usepackage{amsmath}
\usepackage{thmtools}
\usepackage{amssymb}

\usepackage[]{varioref}
\PassOptionsToPackage{hyphens}{url} 

\usepackage{hyperref}
\hypersetup{
      breaklinks=true,
      pdfborder={0 0 0},
      pdfhighlight={/N},
      pdfdisplaydoctitle=true
}

\usepackage{soul}

\usepackage{algorithm}
\newfloat{algorithm}{tp}{lop}
\usepackage[noend]{algpseudocode}
\algnewcommand{\LineComment}[1]{\State \(\triangleright\) #1}

\newcommand{\MPSCD}{\texorpdfstring{\smash{MPS\raise0.6ex\hbox{\scriptsize2}CD}}{MPSCD}}
\newcommand{\tttMPSCD}{\texttt{\smash{MP\!S\raise0.6ex\hbox{\tiny2}CD}}}

\usepackage{cite, paralist}
\usepackage{blindtext}

\makeatletter
\newcommand{\setAlgorithmLine}[1]{\setcounter{ALG@line}{#1}}					
\makeatother

\title{Enhancement of Subjective Content Descriptions\\by using Human Feedback}

\author{
	Magnus Bender\textsuperscript{*},
	Tanya Braun\textsuperscript{\textdagger},
	\mbox{Ralf M\"oller\textsuperscript{*}}, and
	Marcel Gehrke\textsuperscript{*}
}

\address{
\vspace*{0.2cm}
	\textsuperscript{*} Institute of Humanities-Centered Artificial Intelligence, Universität Hamburg,\\Warburgstraße 28, 20354 Hamburg, Germany\\
	\email{\{magnus.bender, ralf.moeller, marcel.gehrke\}@uni-hamburg.de}
	\http{www.uni-hamburg.de/chai}
\vspace*{0.2cm}
	\textsuperscript{\textdagger} Computer Science Department, University of Münster,\\Einsteinstraße 62, 48149 Münster, Germany\\
	\email{tanya.braun@uni-muenster.de}
	\http{www.uni-muenster.de/Informatik.AGBraun}
}

\begin{document}
	\maketitle
	\footnotetext{
		Preprint of an article submitted for consideration in International Journal of Semantic Computing (IJSC)
		© 2024 World Scientific Publishing Company
		\url{https://www.worldscientific.com/ijsc}
	}

	\begin{abstract}
	An agent providing an information retrieval service may work with a corpus of text documents.
	The documents in the corpus may contain annotations such as Subjective Content Descriptions (SCD)---additional data associated with different sentences of the documents.
	Each SCD is associated with multiple sentences of the corpus and has relations among each other.
	The agent uses the SCDs to create its answers in response to queries supplied by users.
	However, the SCD the agent uses might reflect the subjective perspective of another user.
	Hence, answers may be considered faulty by an agent's user, because the SCDs may not exactly match the perceptions of an agent's user.
	A naive and very costly approach would be to ask each user to completely create all the SCD themselves.
	To use existing knowledge, this paper presents ReFrESH, an approach for Relation-preserving Feedback-reliant Enhancement of SCDs by Humans.
	An agent's user can give feedback about faulty answers to the agent.
	This feedback is then used by ReFrESH to update the SCDs incrementally.
	However, human feedback is not always unambiguous.
	Therefore, this paper additionally presents an approach to decide how to incorporate the feedback and when to update the SCDs.
	Altogether, SCDs can be updated with human feedback, allowing users to create even more specific SCDs for their needs.
\end{abstract}


	\section{Introduction}
\label{chapter:introduction}
	A corpus of text documents may contain Subjective Content Descriptions (SCDs)~\cite{MPSCD}, which are additional location-specific data associated with sentences of the text documents.
	SCDs highlight points of interest nearby their location, here the sentence of a document, by providing descriptions, references, or explanations.
	A human or an automated annotation technique, e.g., OpenIE~\cite{OpenIE} or USEM~\cite{USEM}, may create SCDs for a specific corpus.
	In general, when reading a text document, each human gets its own perceptions and views of the text document.
	For example, think about studying for an exam.
	You take the script and start to annotate things you consider crucial with your understanding.
	Consequently, the SCDs added to a corpus by a human are slightly different and \emph{subjective} depending on the particular human.
	Similarly, SCDs estimated with automated annotation techniques depend on the particular technique.

	We assume, we have a corpus associated with SCDs to start with.
	Together, the text documents and the associated SCDs build a model of the corpus.
	An agent~\cite{Russell2021}, which is a rational and autonomous unit acting in a world fulfilling a defined task, uses this model and provides an Information Retrieval (IR) service.
	The IR agent answers queries of users which may be humans or other agents.
	A query is some unseen text to which the user is interested in finding similar and relevant documents from the agent's corpus.
	To answer a query, the IR agent uses the SCDs associated with the corpus and returns documents that are assumed to be similar because they share the same SCDs as the query.
	A user may respond to the IR agent's answer with feedback.
	This feedback can be quite explicit, e.g., the user may report a faulty or not similar document.
	In contrast, the feedback may also be ambiguous, e.g., the user only states to be not satisfied with the IR agent's answer. 

	To generate the answer, the IR agent heavily relies on SCDs.
	However, in most cases, a user of the agent will not create the initial SCDs of the corpus itself---at least not all.
	At this point, there may be a difference between the SCDs used by the agent and the SCDs envisioned by the user.
	In other words, the SCDs used for IR do not necessarily represent the perceptions of the agent's user.
	One possibility to avoid the difference is to force each user to create the initial SCDs of the corpus itself.
	However, having each user annotate the whole corpus with its understanding is not feasible in any realistic setting.
	Therefore, we update the model in case the user determines a mismatch.
	Then, the IR agents changes the SCDs according to the feedback about the mismatch.
	To do so, it needs a technique to incrementally change SCDs.

	A technique to incrementally change SCD-based models is the Feedback-reliant Enhancement of SCDs by Humans (FrESH)~\cite{FrESH}.
	FrESH removes faulty sentences and their SCDs entirely from a corpus.
	However, removing the entire sentence does not solve the first problem this paper addresses:
	A faulty association between an SCD and a sentence needs to be updated based on feedback from a user, but the sentence needs to remain in the corpus. 
	As solution this paper presents \mbox{ReFrESH}, an approach for Relation-preserving Feedback-reliant Enhancement of SCDs by Humans. 
	ReFrESH assigns the sentence with the faulty association to a different and hopefully better fitting SCD.
	The other sentences of the SCD, which are not considered faulty, are also updated.
	Considering all sentences is necessary, as the creation of SCDs consists of multiple steps whereas each step influences the next steps.
	However, influences between steps can not be reproduced retrospectively and thus can not easily be considered by ReFrESH.
	The term relation-preserving emphasizes that relations among SCDs and other sentences associated with SCDs are considered by ReFrESH. 
	A relation, e.g., \emph{homonym}, between two SCDs, one about a river bank and one about a financial institution, is preserved.

	FrESH and ReFrESH can be used with explicit feedback, i.e., if a user reports a faulty sentences or faulty association between an SCD and a sentence the SCDs can be updated. 
	However, more ambiguous feedback can not be consider directly by using FrESH or ReFrESH.
	Thus, the second contribution of this paper is an approach for capturing more ambiguous feedback.
	This feedback is first stored and aggregated before it is used to update the SCDs, e.g., by FrESH or ReFrESH.
	We consider the different kinds of feedback and discuss how to decide which available actions is the best.
	
	The remainder of this paper is structured as follows:
	First, we look at related work.
	Second, we recap the basics of SCDs, the estimation of SCDs and sketch FrESH.
	Afterwards, we formalize the problem of updating a single SCD while preserving relations to other SCDs and sentences and present the solution ReFrESH.
	Next, we describe how to capture different kinds of human feedback and use it to enhance SCDs.
	Finally, we present an evaluation of ReFrESH and conclude afterwards.
	\section{Related Work}
\label{chapter:relwork}
	Before we introduce the preliminaries of SCDs and present ReFrESH, we take a look at related work.
	Incrementally updating or changing already available models has been investigated in different ways, but not with SCDs.
	
	One well-known approach is to pre-train a more general model first and fine-tune it later for a specific task. 
	During fine-tuning the model is trained on new task specific data. 
	A typical example are transformer models like BERT~\cite{BERT-initial} \mbox{or GPT~\cite{GPT-3}.}

	Another possibility is to bring the task specific data in during the computation of the answer.
	In this case, the model is not updated, but to each query some user and case specific data is added before the query is processed by the model.
	Most chat bots like ChatGPT\footnote{\url{https://chat.openai.com/}} or Gemini\footnote{\url{https://gemini.google.com/}, formerly Bard} use this technique.

	Both of these techniques incrementally update a model, like ReFrESH, and work with models based on deep learning.
	Techniques that update a model must be distinguished from techniques that completely remove an item from the model, such as FrESH. 
	There are approaches to remove an item from a model, e.g., for k-Means~\cite{kMeans,AIforget} or linear and logistic regression~\cite{DataDelML}. 
	The common idea is to avoid retraining the model and instead only incrementally change the model by applying an inverse operation that removes a single item of the training data from the model.

	For ReFrESH, we focus on incrementally updating faulty associations between SCDs and sentences, while leaving the corpus unchanged and preserving all sentences.

	\section{Preliminaries}
	This section specifies notations, recaps the basics of SCDs, and the estimation of SCDs.
	Additionally, we sketch FrESH.

	\subsection{Notations}
		First, we formalize our setting of a corpus.
		\begin{compactitem}
			\item
				A word $w_i$ is a basic unit of discrete data from a vocabulary $\mathcal{V} = \{w_1, \dots, w_L\}$, $L \in \mathbb{N}$.
			\item 
				A sentence $s$ is defined as a sequence of words \mbox{$s = (w_1, \dots, w_N)$}, $N \in \mathbb{N}$, where each word $w_i \in s$ is an element of vocabulary $\mathcal{V}$.
				Commonly, a sentence is terminated by punctuation symbols like \enquote{.}, \enquote{!}, or \enquote{?}. 
			\item 
				A document $d$ is defined as a sequence of sentences \mbox{$d = (s^d_1, ..., s^d_M)$}, $M \in \mathbb{N}$.
			\item
				A corpus $\mathcal{D}$ is a set of documents $\{d_1, \dots, d_{D}\}$, $D \in \mathbb{N}$.
			\item 
				An SCD $t$ is a tuple. 
				The tuple contains the SCD's additional data $\mathcal{C}$, i.e., the label $l$ of the SCD and a set $R$ of relations to other SCDs.
				Additionally, each SCD's tuple contains references to the referenced sentences in documents of $\mathcal{D}$, while in the opposite direction a sentence is associated with an SCD.
			\item 
				A sentence associated with an SCD is called SCD window, inspired by a tumbling window moving over the words of a document. 
				Generally, an SCD window might not be equal to a sentence and may be a subsequence of a sentence or the concatenated subsequences of two sentences, too.
				Even though, in this paper, an SCD window always equals a sentence.
			\item 
				For a corpus $\mathcal{D}$ there exists a set $g$ called SCD set containing $K$ associated SCDs
				$
					g(\mathcal{D}) = 
						\left\{
							t_j = \left(
									\mathcal{C}_j,
									\bigcup_{d \in \mathcal{D}}
										\{
											s^d_1, ...., s^d_S
										\}
							\right)
						\right\}^K_{j=1}.
				$
				Given a document $d \in \mathcal{D}$, the term $g(d)$ refers to the set of SCDs associated with sentences from document $d$.
			\item 
				Each word $w_i \in s^d$ is associated with an influence value $I(w_i, s^d)$ representing the relevance of $w_i$ in the sentence $s^d$.
				For example, the closer $w_i$ is positioned to the object of the sentence $s^d$, the higher its corresponding influence value $I(w_i, s^d)$.
				The influence value is chosen according to the task, e.g., distributed binomial, linear, or constant.
		\end{compactitem}

	\subsection{Subjective Content Descriptions}
		SCDs provide additional location-specific data for documents~\cite{MPSCD}.
		The data provided by SCDs may be of various types, like additional definitions or links to knowledge graphs. 

		Kuhr et al.\ use an SCD-word distribution represented by a matrix when working with SCDs~\cite{MPSCD}.
		The SCD-word distribution matrix, in short SCD matrix, can be interpreted as a generative model. 
		A generative model for SCDs is characterized by the assumption that the SCDs generate the words of the documents.
		We assume that each SCD shows a specific distribution of words of the referenced \mbox{sentences in the documents.}

		The SCD matrix $\delta(\mathcal{D})$ models the distributions of words for all SCDs $g(\mathcal{D})$ of a corpus $\mathcal{D}$ and is structured as follows:
		\begin{align*}
			\delta(\mathcal{D}) = 
			\bordermatrix {
				& w_1	& w_2	& w_3	& \cdots	& w_L	\cr
				t_1	& v_{1,1}	& v_{1,2}	& v_{1,3}	& \cdots	& v_{1,L}	\cr
				t_2	& v_{2,1}	& v_{2,2}	& v_{2,3}	& \cdots	& v_{2,L}	\cr
				\vdots& \vdots	& \vdots	& \vdots	& \vdots	& \vdots	\cr
				t_K & v_{K,1}	& v_{K,2}	& v_{K,3}	& \cdots	& v_{K,L}	\cr
			}
		\end{align*}

		The SCD matrix consists of $K$ rows, one for each SCD in $g(\mathcal{D})$.
		Each row contains the word probability distribution for an SCD.
		Therefore, the SCD matrix has $L$ columns, one for each word in the vocabulary of the corpus $\mathcal{D}$.
	
	\subsection{Supervised and UnSupervised Estimator for SCD Matrices}
		The SCD matrix can be estimated in a supervised manner given the set $g(\mathcal{D})$ for a corpus $\mathcal{D}$.
		SEM is described in Algorithm~\ref{alg:scd-matrix}.
		Given a corpus $\mathcal{D}$, the algorithm iterates over each document $d$ in the corpus and the document's SCDs.
		For each associated SCD $t$, the referenced sentences $\{s^d_1, ..., s^d_S\}$ are used to update the SCD matrix.
		Thereby, the row of the matrix representing SCD $t$ gets incremented for each word in each sentence by each word's influence value.
		
		Finally, the SCD matrix needs to be normalized row-wise to meet the requirements of a probability distribution.
		However, the normalization is often skipped because later the cosine similarity is often used with the rows of the matrix and the cosine similarity does a normalization by definition. 

		\begin{algorithm}
			\caption{Supervised Estimator of SCD Matrices $\delta(\mathcal{D})$}
			\begin{algorithmic}[1]
				\Function{SEM}{$\mathcal{D}$, $g(\mathcal{D})$}
					\State \textbf{Input}: Corpus $\mathcal{D}$; Set of SCDs $g(\mathcal{D})$
					\State \textbf{Output}: SCD-word distribution matrix $\delta(\mathcal{D})$
					\State Initialize an $K \times L$ matrix $\delta(\mathcal{D})$ with zeros
					\For {each document $d \in \mathcal{D}$}
						\For {each SCD $t = (\mathcal{C},\{s^d_1,...,s^d_S\}) \in g(d)$}
							\For {$j = 1,...,S$} \Comment{Iterate over sentences}
								\For {each word $w_i \in s^d_j$}
									\State $\delta(\mathcal{D})[t][w_i] \mathrel{+}= I(w_i, s^d_j)$
								\EndFor
							\EndFor
						\EndFor
					\EndFor
					\State \Return $\delta(\mathcal{D})$
				\EndFunction
			\end{algorithmic}
			\label{alg:scd-matrix}
		\end{algorithm}

		Unlike SEM, USEM estimates an SCD matrix $\delta(\mathcal{D})$ without needing the SCD set $g(\mathcal{D})$.
		USEM initially starts by associating each sentence to one unique SCD, which leads to an initial SCD matrix consisting of a row for each sentence in the document's corpus.
		Then, USEM finds the sentences in the corpus that represent the same concept and groups them one by one into an SCD.

		SEM, e.g., along with OpenIE for getting $g(\mathcal{D})$, and USEM are two techniques to get SCDs for a corpus.
		Especially USEM can be used by an IR agent to automatically create initial SCDs for a corpus.
		Afterwards, the agent may use ReFrESH and update the SCDs based in the users feedback.
		
	\subsection{Removing Sentences from Corpus and SCDs}
		A may corpus contain sentences with erroneous content or sentences which are protected by privacy regulations or copyright.
		Such sentences need to be removed entirely from the corpus and the associated SCDs.
		To do so, FrESH~\cite{FrESH} provides a technique for incrementally updating the SCD matrix.
		FrESH does not allow to change SCDs and instead it entirely removes faulty sentences which are associated with an SCD.
		FrESH inverts the operations of SEM, mainly the addition in Line 9 of Algorithm~\ref{alg:scd-matrix}.

		Internally, ReFrESH will need to remove and add sentences of SCDs, too.
		Thus, \textbf{Re}FrESH is the \textbf{Re}lation-preserving update technique for SCD-based models, which builds on top of the much smaller FrESH.
	ReFrESHs uses ideas of FrESH, SEM, and USEM. 	
	Next, we present ReFrESH which provides incremental updates for SCDs while it preserves relations among SCDs and leaves the corpus unchanged.

	\section{Relation-preserving Updates on SCD Matrices}
	\label{sec:refresh}
	This section introduces ReFrESH, the algorithm that allows to update an SCD-based model by correcting faulty associations between sentences and SCDs.
	First, we look at possible relations among SCDs and describe ReFrESH afterwards. 

	\subsection{Relations to Preserve}
		Before we can compose a relation-preserving algorithm, we need to specific the different relations among SCDs.
		An SCD $t_i$ consists of the referenced sentences $\{s_1, ..., s_S\}$, the word distribution $(v_{i, 1}, ..., v_{i,L})$, and its additional data $\mathcal{C}_i$.
		One of the referenced sentences $s_r$ has been marked as faulty and should be removed.
		However, the remaining $S-1$ sentences $\mathcal{S}_{c} = \{s_1, ..., s_S\} \setminus s_r$ are then considered as correct.
		The word distribution will be different after a sentence is removed from an SCD, but can be easily recalculated afterwards.
		All items of the additional data $\mathcal{C}_i$ are related, i.e., each item can be understood as related to its SCD and thus as a relation of the SCD to be preserved.

		Summarized, ReFrESH needs to preserve the relations of the referenced sentences $s_r$ and $\mathcal{S}_{c}$ to the items in $\mathcal{C}_i$.
		In the following, $\mathcal{C}_i$ contains a label $l_i$, e.g, computed by LESS~\cite{LESS}, and a set of relations to other SCDs $R_i$.
		Each Tuple in $R_i$ models a relation of SCDs, e.g., $(t_i, t_j)$ a relation between $t_i$ and $t_j$.

		In general, a sentence may have relations which directly belong to the sentence itself and not to its SCD.
		As ReFrESH only modifies SCDs, relations belonging to a sentence are not effected by ReFrESH.
		ReFrESH uses this by shifting relations from the SCD to the sentences.

\begin{figure*} 
			\centering
			\includegraphics[width=0.99\linewidth]{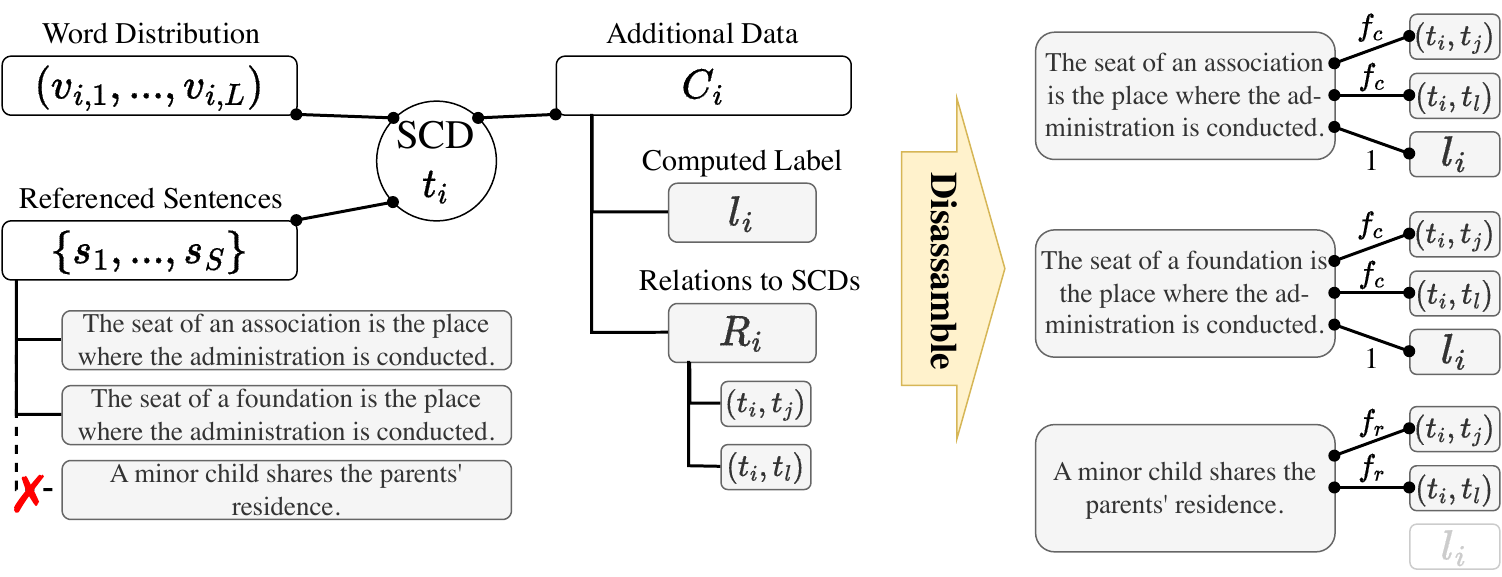}
			\caption{
				Left: An SCD with its referenced sentences, three in this example, its row of the word distribution, and its additional data, e.g., containing a label and two relations to other SCDs.
				The red cross marks the sentence to remove from the SCD.
				Right: The SCD after the disassemble step.
				Each of the three sentences now stand by itself, while the label and the two relations have been reassigned to the individual sentences and given a factor.
			}
			\label{fig:scd-rel-disass}
		\end{figure*}

		On the left hand side of Figure~\ref{fig:scd-rel-disass}, an SCD with the previously described parts is depicted.
		In this example, the SCD references three sentences.
		The sentences with the faulty association is already marked by a red cross.
		The additional data contains a label and two relations of $t_i$ to \mbox{other SCDs, i.e., $t_j$ and $t_l$.}

	\subsection{Four Steps for Updating an SCD}
		ReFrESH needs to preserve the relations between the referenced sentences and the items in the additional data of an SCD.
		Afterwards, the sentence $s_r$ and the sentences $\mathcal{S}_{c}$ shall be associated with different SCDs.

		We assume to have a corpus $\mathcal{D}$ with an SCD matrix $\delta(\mathcal{D})$ and the set of SCDs $g(\mathcal{D})$.
		Together, these three parts build an SCD-based model which is updated by ReFrESH.
		To do so, ReFrESH needs four steps.
		For each step, we provide an algorithm that represents the operations of that step. 
		Before we finally we combine all four steps to get the complete algorithm of ReFrESH in Subsection~\ref{sec:algo-refresh}.

		\subsubsection{Shift Relations to Sentences}
			The input of ReFrESH is a sentence $s_r$ which is falsely associated with an SCD $t_i$. 
			ReFrESH's task is to remove $s_r$ from $t_i$ and to reassign $s_r$ to a better fitting SCD.
			Besides preserving the relations, ReFrESH also needs to consider the impact of $s_r$ on the SCD $t_i$.
			When $s_r$ is associated with $t_i$, the word-distribution of $t_i$ also represents $s_r$.
			Thus, similar sentences to $s_r$ might also be added to $t_i$, just because $s_r$ is associated with $t_i$.
			This is based on the assumption that sentences are added to an SCD one after another and using the word-distribution to measure similarity.
			For example, USEM chooses best matching sentences in a greedy way one after one.
			Also, humans may build their own set of SCDs manually and incrementally (including the use of ReFrESH multiple times in a row). 
			Thus, when $s_r$ is removed, other sentences added because of $s_r$ may also need to be removed from the SCD.

			A comparable scenario arises when clustering data points using a density based clustering algorithm: 
			If a point at a border of a cluster is sufficiently close to another cluster, both clusters may get merged.
			In this case, the point at the border becomes some type of bridge between both clusters and without this point both clusters would not have been merged.
			Hence, if this border point is removed later, both clusters should be separate, too.

			To take this into account, ReFrESH needs to consider all referenced sentences $\{s_1, ..., s_S\}$ of $t_i$.
			To be able to consider every sentences individually, ReFrESH first shifts all the relations to preserve, the ones in $\mathcal{C}_i$, directly to the individual sentence.
			A sentence to remove might not share the topic of an SCD, thus, ReFrESH does not shift the label to $s_r$.
			However, the other sentences $\mathcal{S}_c$, which are considered correct, will share the topic of the SCD.
			Thus, the sentences in $\mathcal{S}_c$ are assigned with the label $l_i$ and the factor $1$ for this relation between sentence and label.

			Next, ReFrESH does the shift for all the relations in $R_i$ and again differentiates between sentences $s_r$ and $\mathcal{S}_c$.
			The relations in $R_i$ are added to $\mathcal{S}_c$ with the factor $f_c$ for correct sentences. 
			Thereby, $f_c$ is defined as $f_c = \frac{S-1}{S}$ where $S$ is the number of referenced sentences of $t_i$.
			In contrast, the relations in $R_i$ are added to $s_r$ with the factor $f_r$ for removed sentences.
			Here, the factor is $f_r = \frac{1}{S}$.
			If some relation already has a factor, both factors are multiplied because the factors express uncertainty.
			
			The relations from $\mathcal{C}_i$ are shifted to the sentences to make sure all relations are preserved.
			The factors make sure that relations are preserved differently for $\mathcal{S}_{c}$ and $s_r$, i.e., if a sentence is considered correct, the relations of the SCD are also more likely to be correct than if the sentence is falsely associated.
			Of course, it might be necessary to change the factor for specific relations and use-cases.
			All sentences and relations are stored in $\mathcal{P}$.

			\begin{algorithm}
				\caption{Step 1) of ReFrESH -- Shift Relations to Sentences}
				\begin{algorithmic}[1]
					\setAlgorithmLine{4}
					\State $\mathcal{P} \leftarrow \emptyset$ \Comment{Sentences with relations to preserve}
					\For {each referenced sentence $s_i \in \{s_1, ..., s_S\}$}
						\If { $s_i = s_r$ } \Comment{Calculate factor}
							\State $f \leftarrow \frac{1}{S}$, $p_i \leftarrow \emptyset$
						\Else \Comment{Preserve label for sentences in $\mathcal{S}_c$}
							\State $f \leftarrow \frac{S-1}{S}$, $p_i \leftarrow \{(1, l_i)\}$
						\EndIf
						\For {each relation to preserve $r_i \in R_i$} 
							\State $p_i \leftarrow p_i \cup \{(f, r_i)\}$ \Comment{Store with factor}
						\EndFor
						\State $\mathcal{P} \leftarrow \mathcal{P} \cup \{(s_i, p_i)\}$ \Comment{Store sentence and rel.}
					\EndFor
				\end{algorithmic}
				\label{alg:refresh-step-1}
			\end{algorithm}

		\subsubsection{Disassemble SCD}
			Coming back to the problem that we do not know why a sentence was added to an SCD.
			ReFrESH can not determine which sentences haven been added because of $s_r$, too.
			The only solution to this problem is to disassemble the entire SCD $t_i$.
			After step 1), all relations to preserve are directly tied to each referenced sentence.
			Thus, the SCD $t_i$ is not needed any more and can be disassembled without loosing important information.
			Afterwards, ReFrESH is able to consider each sentence separately.
			
			Disassembling an SCD means deleting the word distribution $(v_{i,1},...,v_{i,L})$, the $i$-th row, from the SCD matrix and removing $t_i$ from $g(\mathcal{D})$.
			Additionally, the SCDs in relation with $t_i$ are informed that they are now in relation with the referenced sentences of $t_i$. 

			On the right hand side of Figure~\ref{fig:scd-rel-disass}, an example of an disassembled SCD $t_i$ is shown.
			The lowest sentences is $s_r$, in this case the factors are $f_r = \frac{1}{3}$ and $f_c = \frac{2}{3}$.
			The label gets a factor of $1$ for the upper two sentences $\mathcal{S}_c$ and $s_r$ has no label.
			The SCD $t_i$ is temporarily removed completely.

			\begin{algorithm}
				\caption{Step 2) of ReFrESH -- Disassemble SCD}
				\begin{algorithmic}[1]
					\setAlgorithmLine{13}
					\State $g(\mathcal{D}) \leftarrow g(\mathcal{D}) \setminus t_i$
					\State $\delta(\mathcal{D})[i] \leftarrow Nil$ \Comment{Delete $i$-th row}
				\end{algorithmic}
				\label{alg:refresh-step-2}
			\end{algorithm}
	
		\subsubsection{Reassign Sentences to SCDs}
			Now, the previously referenced sentences stored in $\mathcal{P}$ need to be reassigned, i.e., each sentence needs to be associated with a new SCD.
			In this third step, ReFrESH needs to find the best SCD for each sentence.
			This best SCD may be an already known SCD of the corpus or newly composed SCDs while ReFrESH needs to assure that $s_r$ does not get associated with a new SCD.
			A new SCD references only sentences from $\mathcal{S}_c$.
			If $s_r$ gets associated with such new SCD, the association of $s_r$ and the new SCD might be very similar or even equal to the initial SCD $t_i$ before running ReFrESH to remove $s_r$.

			Analogously, in the example about the clustering of data points, all points of both clusters and the border point would be considered again.
			Each point may be added to another cluster or one or more new clusters may be created, while the border point will become an outlier or member of another cluster.

			We still need an algorithm to reassign sentences to SCDs, which is similar to estimating SCDs in an unsupervised way.
			Thus, ReFrESH applies the idea of USEM and uses USEM's greedy method to reassign the sentences with new or known SCDs.
			The idea of USEM is to start by considering each sentences as an SCD with one referenced sentence.
			Afterwards, USEM uses its greedy method and identifies the two most similar SCDs to merge.
			Hence, in the first iteration of USEM two SCDs with one referenced sentence each get merged and become one SCD with two referenced sentences.
			To identify similar SCDs, the cosine similarity is used with the \mbox{SCD matrix' rows.}

			For ReFrESH, the idea is applied as follows:
			First, the word-vector for $s_r$ is calculated (as in Lines 8 and 9 of Alg.~\ref{alg:scd-matrix}) and $s_r$ is added to most similar and already known SCD of the corpus (Lines 17 and 18 in Alg.~\ref{alg:refresh-step-3}).
			For each sentence in $\mathcal{S}_c$ its word-vector is then compared to all rows in the SCD matrix and to the word-vectors of the other sentences (Lines 22 and 23 in Alg.~\ref{alg:refresh-step-3}).
			If a word-vector is most similar to one of the rows in the SCD matrix, and thus to an already known SCD of the corpus, the sentence is added to the SCD as referenced sentence (Lines 25 - 27 in Alg.~\ref{alg:refresh-step-3}).
			In the other case, if a word-vector is most similar to a word-vector of another sentence, both sentences are merged to form a new SCD, which is then added to $g(\mathcal{D})$ and $\delta(\mathcal{D})$ (Lines 29 - 32 in Alg.~\ref{alg:refresh-step-3}).
			This is repeated until all sentences of $\mathcal{S}_c$ are part of an SCD.
			In the end of step 3), ReFrESH recalculates the word distribution of all SCDs to which new referenced sentences have been added.
			The new and modified SCDs are stored in $\mathcal{N}$.

			\begin{algorithm}
				\caption{Step 3) of ReFrESH -- Reassign Sentences to SCDs}
				\begin{algorithmic}[1]
					\setAlgorithmLine{15}
					\State $\mathcal{N} \leftarrow \emptyset$ \Comment{Note changed SCDs}
						\Statex \Comment{First reassign $s_r$}
					\State $j \leftarrow \Call{mostSimilarRow}{\vec{s_r}, \delta(\mathcal{D})}$ 
					\State $\delta(\mathcal{D})[j] \leftarrow \delta(\mathcal{D})[j] + \vec{s_r}$ \Comment{Update matrix}
					\State $t_j \leftarrow (\mathcal{C}_j, \{s_1, ..., s_S\} \cup \{s_r\})$ \Comment{Add $s_r$ to $t_j$}
					\State $\mathcal{N} \leftarrow \mathcal{N} \cup \{(t_j, s_r, p_r)\}$ \Comment{Note that $t_j$ changed}
						\Statex \Comment{Reassign remaining sentences $\mathcal{S}_c$}
					\For {each sentence and rel.\ $(s_i, p_i) \in \mathcal{P} \setminus (s_r, p_r)$}
						\State $j \leftarrow \Call{mostSimilar}{\vec{s_i}, \delta(\mathcal{D})}$
						\State $k \leftarrow \Call{mostSimilar}{\vec{s_i}, \vec{\mathcal{P}} \setminus \vec{s_i}}$
						\If {$\Call{similarity}{j} > \Call{similarity}{k}$} 
							\State $\delta(\mathcal{D})[j] \leftarrow \delta(\mathcal{D})[j] + \vec{s_i}$ \Comment{Update matrix}
							\State $t_j \leftarrow (\mathcal{C}_j, \{s_1, ..., s_S\} \cup \{s_i\})$ \Comment{Add $s_i$ to $t_j$}
							\State $\mathcal{N} \leftarrow \mathcal{N} \cup \{(t_j, s_i, p_i)\}$ 
						\Else \Comment{Build new SCD $t_k$ with $s_i$ and $s_k$}
							\State $\delta(\mathcal{D})[k] \leftarrow \vec{s_i} + \vec{s_k}$ \Comment{Add row to matrix} 
							\State $g(\mathcal{D}) \leftarrow g(\mathcal{D}) \cup (\mathcal{C}_k, \{s_i, s_k\})$ \Comment{Create SCD} 
							\State $\mathcal{P} \leftarrow \mathcal{P} \setminus (s_k, p_k)$ \Comment{$s_k$ already reassigned}
							\State $\mathcal{N} \leftarrow \mathcal{N} \cup \{(t_k, s_i, p_i), (t_k, s_k, p_k)\}$ 
							
						\EndIf
					\EndFor
				\end{algorithmic}
				\label{alg:refresh-step-3}
			\end{algorithm}

		\subsubsection{Propagate new Relations}
			Finally, the SCD-based model contains all sentences again and all SCDs in the model have their word-distribution and set of referenced sentences.
			However, (i) the additional data of all new SCDs is empty and (ii) the SCDs that have received one or more new referenced sentences do not have the relations of their new sentences.
			Algorithm~\ref{alg:refresh-step-4} iterates through all modified SCDs in $\mathcal{N}$ including the sentences and relations and addresses both cases.
	
			In the case of (i), the new SCD does not have any relations itself.
			Furthermore, the relations of the referenced sentences of this SCD are all the same, as all sentences originate from the same disassembled SCD.
			Thus, the relations can be shifted back from the sentences to the SCD including the factors.
			The factors outline some uncertainties, as relations may not originally originate from the SCD and its sentences.
			Finally, a new label for the SCD is calculated by LESS~\cite{LESS}.

			Otherwise, case (ii), we assume that $x$ new sentences have been added to an SCD with previously $S$ sentences.
			In contrast to (i), the relations stay with the sentences and are also propagated to the SCD.
			The relations from the sentences are added to the SCD's additional data and each factor is multiplied by $\frac{x}{S+x}$.
			By using this factor, each relation is weighted depending the ratio it has among the referenced sentences of the SCD.
			The label of the SCD will not be changed, but labels from the sentences are added like a relation including the factor.
			The factors used with the relations by ReFrESH are slightly inspired by weighted model counting~\cite{WeightedModelCounting}.	

			Finally, the original association of $s_r$ with $t_i$ has been removed and all former referenced sentences of $t_i$ have been reassigned to a new and better fitting SCD.
			All relations have been preserved and propagated to other SCDs, too.

			\begin{algorithm}
				\caption{Step 4) of ReFrESH -- Propagate new Relations}
				\begin{algorithmic}[1]
					\setAlgorithmLine{32}
					\For {each SCD, sentence, and rel.\ $(t_i, s_i, p_i) \in \mathcal{N}$}
						\If {$\mathcal{C}_i = \emptyset$} \Comment{New SCD}
							\State $\mathcal{C}_i \leftarrow \mathcal{C}_i \cup p_i$
							\State $l_i = \Call{LESS}{t_i}$
						\Else 
							\For {each factor and relation $(f_i, r_i) \in p_i$}
								\State $\mathcal{C}_i \leftarrow \mathcal{C}_i \cup \left( \frac{x}{S-x} \cdot f_i, r_i \right)$
							\EndFor
						\EndIf
					\EndFor
				\end{algorithmic}
				\label{alg:refresh-step-4}
			\end{algorithm}

		Generally, it is possible to slightly adapt ReFrESH and remove and reassign more than one sentence from an SCD, then the factors need to be adapted.
		Additionally, it would be possible to ask the human users for an advice to which SCD $s_r$ should be reassigned.

	\subsection{Algorithm ReFrESH}
		\label{sec:algo-refresh}
		Based on the previous subsection presenting the four steps of ReFrESH, the complete algorithm is sketched in Algorithm~\ref{alg:refresh-all}.
		ReFrESH follows the four steps using the operations presented for each step and returns the updated SCD-based model as triple $(\mathcal{D}, \delta(\mathcal{D}), g(\mathcal{D}))$.
		In addition, the input contains the sentence to remove $s_r$ and its SCD $t_i$.
		SCD $t_i$ might be omitted, then all SCDs in $g(\mathcal{D})$ must be searched for the SCD associated with $s_r$.

		\begin{algorithm}
			\caption{Relation-preserving Feedback-reliant Enhancement of SCDs}
			\begin{algorithmic}[1]
				\Function{ReFrESH}{$(\mathcal{D}, \delta(\mathcal{D}), g(\mathcal{D}))$, $s_r$, $t_i$} 
					\State \textbf{Input}: SCD-based model $(\mathcal{D}, \delta(\mathcal{D}), g(\mathcal{D}))$,\\
						\hspace*{1.65cm} sentence to remove $s_r$, and associated SCD $t_i$
					\State \textbf{Output}: Updated model $(\mathcal{D}, \delta(\mathcal{D}), g(\mathcal{D}))$
					\setAlgorithmLine{4}
					\State \Comment{\textbf{Step 1)} Shift Relations to Sentences \hphantom{aaaaaaaaaaaaaaaaaaaaaaaaaaaaaa}}
					\Statex $\vdots$ \Comment{Operations of Algorithm~\ref{alg:refresh-step-1}}
					\setAlgorithmLine{13}
					\State \Comment{\textbf{Step 2)} Disassemble SCD \hphantom{aaaaaaaaaaaaaaaaaaaaaaaaaaaaaaaaaaaaaaa}\,}
					\Statex $\vdots$ \Comment{Operations of Algorithm~\ref{alg:refresh-step-2}}
					\setAlgorithmLine{15}
					\State \Comment{\textbf{Step 3)} Reassign Sentences to SCDs \hphantom{aaaaaaaaaaaaaaaaaaaaaaaaaaaaaa}}
					\Statex $\vdots$ \Comment{Operations of Algorithm~\ref{alg:refresh-step-3}}
					\setAlgorithmLine{32}
					\State \Comment{\textbf{Step 4)} Propagate new Relations \hphantom{aaaaaaaaaaaaaaaaaaaaaaaaaaaaaaaa}\,}
					\Statex $\vdots$ \Comment{Operations of Algorithm~\ref{alg:refresh-step-4}}  
					\setAlgorithmLine{39}
					\State \Return $(\mathcal{D}, \delta(\mathcal{D}), g(\mathcal{D}))$
				\EndFunction
			\end{algorithmic}
			\label{alg:refresh-all}
		\end{algorithm}

	We have now proposed the algorithm of ReFrESH with four steps.
	Next, we describe how to capture different kinds of human feedback and use it to enhance SCDs.
	Afterwards, we present the evaluation of ReFrESH, by describing and discussing the dataset, workflow, and metrics along with the results.
	\section{SCD Enhancement using Feedback}
	In this section, we discuss how IR agents working with SCDs use feedback from human users to enhance their SCDs and thereby improve IR for their users.
	To get into the topic, we first describe an IR agent using SCDs for IR, i.e., an SCD-based IR agent (SCD-IRa).
	An SCD-IRa perceives human feedback in a variety of ways and must decide whether to take an action or gather feedback for a later action.
	Hence, we present possible actions an SCD-IRa can conduct to maintain its SCDs.
	Finally, we present and discuss an algorithm using the gathered feedback to conduct actions. 

	The goal of this section is to depict the different ways an SCD-IRa percepts feedback and the possible actions to perform on the SCDs to incorporate the feedback.
	It is explicitly not about the human oriented part of human-aware agents, but about presenting the required approaches needed to assemble such an agent.

	\subsection{Intelligent Agents}
		An agent~\cite{Russell2021} is a rational and autonomous unit acting in a world fulfilling a defined task.
		The world the agent exists in provides the environment.
		The agent is able to perceive this environment through sensors.
		Additionally, the agent is able to conduct actions through actuators and thus change the environment.
		Generally, the available sensors and actions depend on the actual agent and its task.
		
		Internally, the agent is powered by decision, planning and learning processes, whereas some agents may only use a subset of these processes.
		The agent uses its sensors to model the current environment in an internal state.
		Based on this state, the agent needs to decide which action to do next.
		Normally, the agent needs to select the \emph{best} action from a set of possible actions.
		Such a selection requires knowledge about the consequences of each action and also of other influences the environment is exposed to, e.g., other agents acting in the same environment.
		To select the \emph{best} action, the agent needs to calculate a score for each action considering the environment after conducting this action.
		This score represents the utility of each action, with the agent's goals in the background.
		We assume that the goals of the agent represent an abstract task the agent has to fulfill.
		Based on the goals, the agent is able to plan a sequence of actions bringing it near the goals.

	\subsection{SCD-based IR Agent}
		Let us consider the following abstract use-case for an SCD-IRa:
		A human user brings a corpus of text documents to work with.
		This corpus represents the general field of interest for the user.
		The user might either be a human or another agent.
		The user sends queries about the corpus to the SCD-IRa and SCD-IRa responds to those queries using SCDs.
		To do so, SCD-IRa uses SCDs and an SCD matrix for the user supplied corpus.

		The SCDs may be created using SEM or USEM, but they can also be added manually by the user, other humans, or agents.
		Furthermore, the SCDs are not limited to representing clusters of similar sentences.
		Additionally, complementary SCDs\cite{cSCD,cSCD-more} and SCDs with weighted inter-SCD relations, e.g., created by the propagate step of ReFrESH, may be associated with the corpus.

		SCD-IRa maintains the corpus associated with SCDs and the SCD matrix to answer queries.
		We assume a query consists of one or multiple sentences to which relevant documents should be retrieved from the corpus.
		Relevant documents may be similar documents sharing the same SCD and documents associated with a related SCD like a complementary SCD.
		For each SCD, for example, similar and complementary, their referenced sentences are returned as response, whereas each sentence highlights a relevant part of a document.
		
		The goal of SCD-IRa is to satisfy the information need of its users by providing IR responses.
		A user's satisfaction with an SCD-IRa response depends on the correspondence between the SCDs used by SCD-IRa and the SCDs envisioned by the user.
		In other words, if the SCDs used by SCD-IRa do not represent the perceptions of the agent's user, the user will not be satisfied with the response.
		Hence, SCD-IRa needs to gather feedback and update the SCDs to better match the user's perceptions, such that it fulfills its goal.
		
		Next, we describe the possibilities to percept feedback.

	\subsection{Perception of Human Feedback}
		We distinguish between two types of feedback.
		Explicit feedback is directly sent as feedback by the user to the agent and reports a concrete issue to take care of:
		\begin{description}
			\item[Faulty Sentence]
				A user reports a faulty sentence in a text document of the corpus. 
				Faulty means errors in the content or privacy or copyright related issues.
				SCD-IRa has to remove the sentence and its references to SCDs.
				If an SCD only references the faulty sentence, this SCDs has to be removed as well.
			\item[Faulty Association]
				A user reports a faulty association of a sentence with an SCD.
				In this case, the SCD itself as well as the content of the sentence are valid, but the association of both is incorrect.
				SCD-IRa has to remove this association, i.e, the referenced sentence from the SCD.
			\item[Revert Changes]  
				After reporting feedback, a user may notice that the IR results have not improved.
				Furthermore, the user may notice that the update not improved the model or notice that there has not been a fault.
				Then, the user may conclude to revert the last update and report to SCD-IRa to revert the a model update.
		\end{description}
		
		In contrast to explicit feedback, implicit or ambiguous feedback is not sent directly to the agent as feedback.
		Rather, implicit feedback results from the use of the agent by the user.
		The difficulties are that SCD-IRa uses a broad range of SCDs, e.g., SCD calculated by SEM and USEM, SCDs from human annotators, and SCDs generated by automatic annotation agents.
		Additionally, the SCDs may represent relations like similarity or complementarity.
		Hence, there are many possible reasons why a user is not satisfied with the IR response while SCD-IRa needs to decide whether and how the SCDs need to be updated.

		SCD-IRa captures implicit feedback by analyzing the actions a user does next after getting an IR response.
		Each of the following actions allow SCD-IRa to draw conclusions about the user's satisfaction with the IR response:
		\begin{description}
			\item[New Query]
				The user directly performs a new IR request with a different, possibly similar, query.
				Thus, we assume the user did not see any relevant information in the IR response.
				However, SCD-IRa is not able to determine if the user decided to concretize the query or to move on to the next topic.
			\item[Select SCD(s)] 
				The user selects one SCD form the IR response and requests more information about this SCD.
				Hence, the user assumes a relevance in this SCD and SCD-IRa can strengthen the association of the SCD and the query terms.
				If the user selects another second SCD directly afterwards, the second SCD might actually be even more relevant.
			\item[Select Sentence(s)]
				The user selects a sentence or document form the IR response.
				Thus, this sentence satisfies the user's information need and the user assumes a relevance of the document with this sentence.
				SCD-IRa can strengthen the association of the SCD and the selected sentence.
				Again, if the user selects another second sentence directly afterwards, the second sentence might actually be even more relevant.
			\item[Add Data]
				The user may add new data, i.e., send new documents, manually add an SCDs, or add a relation.
				In this case, the user misses some information or links in the corpus.
				For SCD-IRa there is an explicit part, i.e., add this data, which can be done easily.
				Additionally, there is the implicit information that the user misses information in the IR response.
		\end{description}

		Next, we specific the approaches available to SCD-IRa for updating its SCDs.

	\subsection{Updates of SCDs}
		SCD-IRa chooses between several approaches to update its SCD-based model.
		First, there are approaches which directly change the SCDs:
		\begin{description}
			\item[FrESH \normalfont \cite{FrESH}]
				Deletes a faulty sentence entirely form the SCD-based model, i.e., the corpus, SCDs, and SCD matrix. 
			\item[ReFrESH] [Section \ref{sec:refresh}]
				Removes a faulty association between an SCD and a sentences and reassign the sentence to a better fitting SCD.
			\item[Restore]
				Reverts the SCD-based model to a previous state.
				Therefore, SCD-IRa needs to store previous versions of its SCD-based model.
		\end{description}

		Second, there are additional lightweight approaches which are used to gather information from implicit feedback.
		\begin{description}
			\item[Response Count]
				This counter shows how often the item has been presented to a user.
				Each time an SCD or a referenced sentence of an SCD is send to a user as part of an IR response, SCD-IRa increments the \emph{response counter} for this item.
			\item[Select Count] 
				This counter shows how often the item has been relevant for a user.
				Each time the user selects a sentence or an SCD from an IR response, SCD-IRa increments the \emph{select counter} for this item.
		\end{description} 
		Using these counters, SCD-IRa gathers information over time.
		In case, SCD-IRa is certain enough, it applies an appropriate approaches to update its SCDs.

		In our understating, enhancement based on feedback does not cover updates of SCDs requested manually by a user.
		In other words, a user manually adding an SCD, a document, or relations is not feedback to consider here.
		For a new document, USEM or SEM may be used to calculated initial SCDs, while manually specified SCDs or relations can simply be added by SCD-IRa to its SCD-based model.

	\subsection{Incorporation of Human Feedback}
		SCD-IRa percepts explicit as well as implicit human feedback and has multiple approaches to update its SCDs.
		The following three steps handle completely the explicit feedback, i.e, SCD-IRa applies
		\begin{enumerate}[(i)]
			\item FrESH in the case of faulty sentences,
			\item ReFrESH to fix faulty associations, and
			\item runs a restore of a previous version to revert changes on the SCDs.
		\end{enumerate}

		Having dealt with the explicit feedback, for the implicit feedback, SCD-IRa can not directly update its SCDs.
		Thus, the counters are used to gather information before the SCDs are updated.
		In doing so, several implicit feedbacks are aggregated before an update is conducted.
		Hence, when
		\begin{enumerate}[(i)]
			\item
				perceiving a new query, no counters are updated because the feedback it too vague,
			\item
				selecting an SCD, the response counter gets incremented for all items in the response and the select counter for the selected SCD, 
			\item 
				selecting a sentence, the response counter gets incremented for all items in the response and the select counter the selected sentence and its associated SCD, and
			\item 
				adding data, the response counter gets incremented for all items in the response.
		\end{enumerate}
		Afterwards, the SCDs are updated by Implicit Feedback Incorporation (IFI) presented in Algorithm~\ref{alg:scd-enhance}.
		IFI analyzes the counters and runs FrESH or ReFrESH if necessary.
	
		\begin{algorithm}
			\caption{Implicit Feedback Incorporation for SCD-based IR agents}
			\begin{algorithmic}[1]
				\Function{EnhanceSCDs}{$\mathcal{M}$, $\theta_{\mathrm{ReFrESH}}$, $\theta_{\mathrm{FrESH}}$} 
					\State \textbf{Input}: SCD-based model $\mathcal{M} = (\mathcal{D}, \delta(\mathcal{D}), g(\mathcal{D}))$,\\
						\hspace*{1.65cm} thresholds $\theta_{\mathrm{ReFrESH}}$ and $\theta_{\mathrm{FrESH}}$
					\State \textbf{Output}: Updated model $\mathcal{M}$
					\Statex
					\For {each document $d \in \mathcal{D}$} \Comment{Iterate over items in model $\mathcal{M}$}
						\For {each SCD $t = (\mathcal{C},\{s_1,...,s_S\}) \in g(d)$}
							\For {$j = 1,...,S$} 
								\State $(rc_t, sc_t) \leftarrow \Call{GetResponseAndSelectCounter}{t}$
								\State $(rc_{s_j}, sc_{s_j}) \leftarrow \Call{GetResponseAndSelectCounter}{s_j}$

								\Statex
								\If{$rc_t \geq \theta_{\mathrm{ReFrESH}} \land rc_{s_j} \geq \theta_{\mathrm{ReFrESH}}$}
									\If{$\frac{sc_t}{rc_t} \cdot \frac{1}{S \cdot 2} > \frac{sc_{s_j}}{rc_{s_j}}$} 
										\State $\mathcal{M} \leftarrow \Call{ReFrESH}{\mathcal{M}, s_j, t}$
											\Comment{ReFrESH, remove $s_j$ from $t$}
										\State \Call{ResetCounter}{$rc_t, sc_t, rc_{s_j}, sc_{s_j}$}
									\EndIf
								\EndIf 

								\Statex
								\If {$rc_{s_j} \geq \theta_{\mathrm{FrESH}}$} 
									\If{$\frac{1}{\theta_{\mathrm{FrESH}}} > \frac{sc_{s_j}}{rc_{s_j}}$} 
										\State $\mathcal{M} \leftarrow \Call{FrESH}{\delta(\mathcal{D}), \{ s_j \}}$
											\Comment{FrESH, remove $s_j$}
										\State \Call{ResetCounter}{$rc_{s_j}, sc_{s_j}$}
									\EndIf
								\EndIf
							\EndFor
						\EndFor
					\EndFor
					\Statex
					\State \Return $\mathcal{M}$
				\EndFunction
			\end{algorithmic}
			\label{alg:scd-enhance}
		\end{algorithm}

		IFI works similar to pruning.
		It checks the counters incremented by the user's actions and updates the SCDs using FrESH or ReFrESH.
		It starts by iterating over each document, SCD, and referenced sentence.
		For each sentences, the algorithm fetches the counter values, i.e., the \emph{response counter} and \emph{select counter} for the sentences and its SCD.

		In Line 10, IFI uses the threshold $\theta_{\mathrm{ReFrESH}}$ to test if the current SCD $t$ and sentence $s_j$ has been contained in an IR response sufficiently often.
		With a threshold of, e.g., $\theta_{\mathrm{ReFrESH}} = 10$, an update is conducted at the earliest after sending $10$ IR responses containing $t$ and $s_j$.
		Next, in Line 11, IFI checks if ReFrESH should be used to remove the association between sentence $s_j$  and SCD $t$.
		$\frac{sc_t}{rc_t}$ is the observed probability for selecting SCD $t$ when sent to a user.
		The number of referenced sentences of $t$ is $S$, s.t., $\frac{1}{S}$ represents the probability of randomly selecting one referenced sentence.
		Overall, IFI uses $\frac{sc_t}{rc_t} \cdot \frac{1}{S \cdot 2}$, which represents half of the probability of randomly selecting one referenced sentence of $t$.
		This probability is compared to $\frac{sc_{s_j}}{rc_{s_j}}$, the observed probability of selecting the currently considered sentence $s_j$ after it is sent to a user.
		If half of the probability when selecting randomly is larger than the observed probability for a sentence, ReFrESH is used to remove the association in Line 12.

		Also for FrESH, a threshold is first used to test whether the currently considered sentence $s_j$ has been contained in an IR response sufficiently often (Line 14).
		A reasonable choice could be, e.g., $\theta_{\mathrm{FrESH}} = 100$, because FrESH entirely deletes a sentence and thus should be used with caution and sufficiently more feedback.
		Again, $\frac{sc_{s_j}}{rc_{s_j}}$ represents the probability of selecting the sentence $s_j$ after it is sent to a user.
		If $\frac1{\theta_{\mathrm{FrESH}}}$ is larger than  $\frac{sc_{s_j}}{rc_{s_j}}$, FrESH is used to remove the sentence $s_j$ in Line 16.
		In fact, this means that $s_j$ has been send to a user at least $\theta_{\mathrm{FrESH}}$ times, while the user never selected it.

		Summarized, SCD-IRa uses its perceptions as feedback and runs FrESH or ReFrESH directly for explicit feedback or gathers counts based on implicit feedback.
		The counts are then used by IFI to run FrESH or ReFrESH if reasonable.

	\subsection{Discussion}
		Our approach for incorporating feedback in SCD-IRa assumes that there is only one user.
		Multiple users will have multiple perceptions and views of the corpus and SCDs, which can lead to conflicting feedback.
		Thus, SCD-IRa with multiple users may maintain a model per user or implement a voting mechanisms for explicit feedback.

		Generally, we combine reinforcement learning and weighted model counting~\cite{WeightedModelCounting} as inspiration for our approach.
		We count the user's interactions with SCDs and sentences in the IR responses and interpret them as feedback.
		The counts are then used to update the SCDs following the goal to keep the good SCDs and sentences with many interactions and remove the ones with less interactions.
		The counts are used to model probabilities, e.g., we compare the observed probability for a sentence to be associated with an SCD with the random one.

		Unsupervised or intrinsic approaches often comprise the risk of automatically choosing a wrong action.
		Additionally, ReFrESH may not be able to find a better fitting SCD to which a sentence can be reassigned.
		In the end, this may lead to a situation where the the SCDs of SCD-IRa are better before an update than afterwards.
		However, SCD-IRa is still used by humans, s.t., a user will notice issues with an update and can use the \emph{revert changes} feedback to restore an older and better SCD-based model.

		SCD-IRa is an abstract concept, for a concrete use-case, it would require more adjustment.
		Such adjustment include, e.g., choosing values for $\theta_{\mathrm{FrESH}}$ and $\theta_{\mathrm{ReFrESH}}$ as well as evaluations involving humans.

		Next, we present the evaluation of ReFrESH and conclude afterwards.
	\section{Evaluation}
\label{chapter:evaluation}
	After we have introduced ReFrESH in Section~\ref{sec:refresh}, we present an evaluation.
	First, we introduce the corpus.
	Afterwards, we describe workflow of the evaluation and the used metrics.
	Finally, we present the results and discuss the performance of ReFrESH.
	
	\subsection{Dataset}
		In this evaluation we use the Bürgerliches Gesetzbuch (BGB)\footnote{\url{https://www.gesetze-im-internet.de/bgb/}, English translation \url{https://www.gesetze-im-internet.de/englisch_bgb/}}, the civil code of Germany, in German language as corpus.
		The BGB is freely available and can be downloaded as XML file.
		Therefore, it is easily parsable and processable.
		As the corpus is a law text it consists of correct language, i.e., punctuation and spelling follow the orthographic rules.
		Thus, less preprocessing and no data cleaning is needed. 
		Furthermore, the words used in text documents have a clear meaning and mostly the same words are used \mbox{instead of using synonyms.}

		We use the first part of the BGB, the so called \enquote{General Part}:
		The entire corpus consists of $228$ law paragraphs and overall $854$ sentences which are used as SCD windows.
		Each law paragraph contains between $1$ and $40$ sentences with an average of $3.78$ sentences.
		The vocabulary consists of $1\,436$ words, where each sentence is between $1$ and $20$ words long with an average of $7.11$ words.

	\subsection{Workflow and Implementation}
		ReFrESH is implemented using Python and runs inside a Docker container.
		The implementation uses the libraries Gensim\footnote{\url{https://radimrehurek.com/gensim/}}, NumPy\footnote{\url{https://numpy.org/}}, and NLTK\footnote{\url{https://www.nltk.org/}}. 
		The evaluation is performed on a machine featuring 8 Intel 6248 cores at 2.50GHz (up to 3.90 GHz) and 16GB RAM.
		We run the following workflow to evaluate ReFrESH:
		\begin{enumerate}[(i)]
			\item 
				Randomly choose a set of pairs of sentences which do not share a similar concept, the set contains around one eighth of all the sentences of the corpus.
				Each pair of sentences is associated with the same SCD and then acts as faulty associations of SCD and sentences.
			\item Estimate a SCD-based model which contains the faulty associations chosen in (i):
				Use USEM with the greedy method and estimate the \emph{faulty} SCD matrix $\delta_f(\mathcal{D})$.
				We add a step to USEM after the initial SCD matrix is created.
				This step groups each pair of sentences from (i) into the same SCD, which leads to faulty associations in the model.  
				Afterwards, USEM continues normally, i.e., finds similar sentences and groups them into SCDs.
			\item
				Run ReFrESH to update $\delta_f(\mathcal{D})$ and remove all the faulty associations initiated by the pairs of sentences from (i).
				Meanwhile, keep a copy of $\delta_f(\mathcal{D})$ and create the new \emph{refresh}ed SCD matrix $\delta_r(\mathcal{D})$, where all the faulty associations have been removed. 
			\item Create a baseline model which represents the correct model for the corpus $\mathcal{D}$.
				Estimate the \emph{baseline} SCD matrix $\delta_b(\mathcal{D})$ using USEM without the additional step.
			\item
				Compare the differences between the three models, i.e., the matrices $\delta_f(\mathcal{D})$, $\delta_r(\mathcal{D})$, and $\delta_b(\mathcal{D})$.
		\end{enumerate}

		This workflow mainly focusses on evaluating step two (disassemble) and step three (reassign).
		The reassignment of sentences to new and better SCDs is the crucial and approximative part of ReFrESH.
		It is important to maintain the relations of the SCDs and sentences, but their treatment is fixed by the algorithm and not approximate.

	\subsection{Metrics}
		Based on the three matrices $\delta_f(\mathcal{D})$, $\delta_r(\mathcal{D})$, and $\delta_b(\mathcal{D})$ we need to evaluate the performance of ReFrESH.
		The main idea is that the distributions of our baseline $\delta_b(\mathcal{D})$ and the refreshed $\delta_r(\mathcal{D})$ should be identical.
		For $\delta_r(\mathcal{D})$ first some faulty associations have been added and removed afterwards by ReFrESH, while $\delta_b(\mathcal{D})$ is trained straightforward.
		Thus, we need to measure the difference between matrices of distributions.
		
		Using the Hellinger distance~\cite{Hellinger}, the distance between two matrices $P$ and $Q$ can be calculated row-wise by:
		\[
			h_i(P, Q) =
				\frac{1}{\sqrt{2}}
				\sqrt{
					\sum_{j=1}^L \left(
						\sqrt{
							P[i][j]
						}
						-
						\sqrt{
							Q[i][j]
						}
					\right)^2
				}
		\]
		The resulting distance vector $H(P, Q)$ contains in each row $h_i(P, Q)$ the distances between the matrix' rows.
		Based on this distance vector, we calculate two metrics:

		First, the proportion of differences, which is the proportion of rows in $H$ which are not equal to zero.
		It shows how many SCDs are different between two SCD matrices.
		Second, the average Hellinger distance, it considers only the non equal rows in $H$ and represents the average difference in $H$.
		It shows how similar the SCDs of two SCD matrices are. 

		A technical note:
		The distances can not be calculate between two SCD matrices directly, because the row numbers of an SCD matrix might change between multiple runs of USEM or ReFrESH.
		Thus, we first create intermediate matrices which use a globally equal window number and calculate the distances on these intermediate matrices.

		Another metric for the evaluation of ReFrESH could be the runtime.
		The difference of the duration running ReFrESH and estimating an new SCD matrix using USEM could be calculated.
		In this evaluation, this should be the seconds needed for generating the baseline matrix $\delta_b(\mathcal{D})$ minus the seconds needed for the refreshed matrix $\delta_r(\mathcal{D})$.
		However, such a difference does not represent the common use-case of ReFrESH.
		Commonly, an SCD-based model would exists and a user would request to correct one faulty association. 
		Then, ReFrESH would be run for a single update.

		Comparing the time needed by USEM, for estimating an entire SCD matrix, and ReFrESH, for updating a single SCD, is not a fair comparison.
		Especially, as USEM's runtime depends on the number of available SCDs.
		Adding a factor, i.e., calculating the runtime per SCD, is also not fair because USEM is intended to process a batch of sentences while ReFrESH does a complex operation with a single SCD.  

		Therefore, the implementation of ReFrESH is not optimized to process batches of faulty associations, as the common use-case is to remove one faulty association and wait for new user inputs.
		Consequently, we do not evaluate the runtime. 

	\subsection{Results}
	\label{sec:results}
\begin{figure*} 
			\centering
			\includegraphics[width=0.99\linewidth]{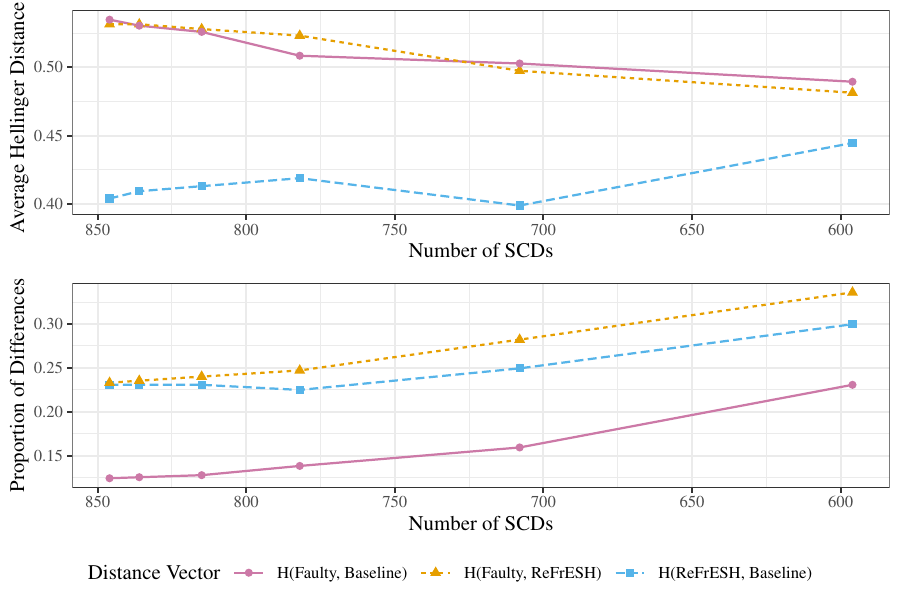}
			\caption{
				Proportion of different rows and average Hellinger distance value for multiple hyperparameters of USEM resulting in different numbers of SCDs.
				For each number of SCDs, three distances, each for one of the distance vectors between the three generated matrices, are shown.
			}
			\label{fig:lines}
		\end{figure*}

		In this section, we present the results gained using ReFrESH and the previously described workflow.
		In the upper graph of Figure~\ref{fig:lines}, the average Hellinger distance is shown for different numbers of SCDs.
		Each number of SCDs represents one run of the workflow and thus three matrices \emph{faulty}, \emph{baseline}, and \emph{refresh}ed.
		Using the three matrices we calculate three Hellinger distance vectors $H(Faulty, Baseline)$, $H(Faulty, ReFrESH)$, and $H(ReFrESH, Baseline)$ and for each vector both metrics.

		The performance of ReFrESH is especially shown by the dashed blue line of $H(ReFrESH, Baseline)$.
		A perfect deletion of faulty associations from the SCD matrix would result in a distance of zero.
		In this case, the distance is greater than zero, but well below the other two lines.
		The solid purple line shows $H(Faulty, Baseline)$, which can be interpreted as the \emph{error} an SCD-based model would have without ReFrESH, because a faulty matrix with all the faulty associations is compared to the baseline.
		Since the dashed blue line is below the solid purple line, ReFrESH reduces the \emph{error} of the model by removing faulty associations.

		In the lower graph of Figure~\ref{fig:lines}, the proportions of different rows are shown---again for different numbers of SCDs and based on three Hellinger distance vectors.
		The two dashed lines represent a distance to $ReFrESH$ and are above the solid purple line of $H(Faulty, Baseline)$.
		A smaller amount of different rows in $H(Faulty, Baseline)$ may be explained by the fact that both models use USEM.
		In contrast, ReFrESH is a different technique and reassigns sentences to other SCDs which in total affects more SCDs.

		\begin{figure} 
			\centering
			\includegraphics[width=0.55\linewidth]{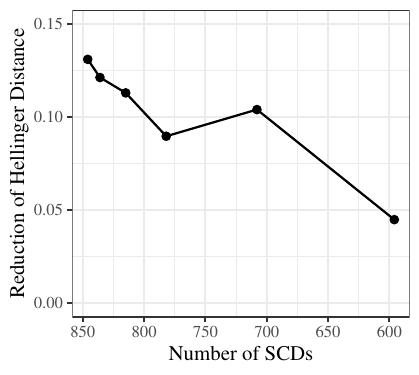}
			\caption{
				Reduction of the Hellinger distance when running ReFrESH on the faulty model and comparing to the baseline, i.e., the amount of correction done by ReFrESH.
			}
			\label{fig:reduction}
		\end{figure}

		In Figure~\ref{fig:reduction}, the reduction of the Hellinger distance by running ReFrESH is shown.
		It shows the average difference of $H(Faulty, Baseline)$ and $H(ReFrESH, Baseline)$.
		In other words, the space between the dashed blue and solid purple line in the upper part of Figure~\ref{fig:lines}.
		Hence, the value can be seen as an improvement of the model when using ReFrESH.

		At first glance, the improvement might be a bit small.
		However, ReFrESH leads to matrices with more different rows but with smaller distances of each row.
		In comparison, the baseline and the faulty model share some identical rows, with each distance value being significantly larger.
		ReFrESH does the reassignment of the sentences to SCDs with the goal of finding a better matching SCD.
		To do so, ReFrESH needs to change many SCDs with the goal of getting a slightly changed but better model.
		Summarized, ReFrESH provides a good performance for refreshing an SCD-based model based on, e.g., human, feedback.

	\section{Conclusion}
\label{chapter:summary}
	This paper first introduces ReFrESH, an approach consisting of four steps for incorporating feedback in SCD-based models.
	In the second part, this paper presents an approach for SCD-based IR agents to utilize human feedback for improving their SCDs.
	SCDs are additional information associated with corpora of text documents and highlight points of interest nearby their location.
	In general, when reading a text document, each human gets its own perceptions and views of the text document.
	Hence, SCDs might by slightly different depending on their origin, i.e., different humans or different automated annotation techniques might yield different SCDs given the same text document.
	If SCDs are used by an IR agent, a user may consider some answers of the agent faulty and respond with feedback to the agent.
	However, this feedback is not always explicit, e.g., the agent may not simply apply ReFrESH if it is not known which association is faulty.
	Therefore, besides ReFrESH in the first part, in the second part we present a decision approach for the agent to decide what to do with the feedback and when to use ReFrESH or FrESH.

	ReFrESH allows incremental updates of SCD-based models based on user feedback and avoids the need for each user to create their own SCDs for each corpus from scratch.
	The first step of ReFrESH shifts all relations among SCDs to the sentences to ensure that relations between SCDs are preserved during the update.
	Second, the SCD to be updated is disassembled before each sentence is reassigned to a better fitting SCD in the third step.
	Finally, the preserved relations are propagated back from the sentences to the SCDs.
	
	Overall, the evaluation shows that ReFrESH works well and provides a powerful technique to update SCD-based models based on human feedback.
	Using ReFrESH, faulty association between sentences and SCDs can be removed and the sentences get associated with new and better fitting SCDs.
	The evaluation focusses on step two (disassemble) and step thee (reassign) because the crucial and approximative part of ReFrESH is the reassignment of an SCD.

	In the second part, IFI is used to analyze the implicit feedback gathered from users and decides whether to run ReFrESH and FrESH to improve the SCDs.
	In the field of SCDs, both contributions of this paper provide an important step for using SCD-based models in IR agents.
	ReFrESH provides an update approach for SCDs and IFI helps the IR agent to run updates based on implicit feedback.
	An SCD-based IR agent is now able to incorporate feedback from its users and decide how to update its model based on this feedback.

	Using ReFrESH on SCDs may not always result in improved SCDs.
	Currently, the IR agent offers the option to revert the entire model to a previous version.
	However, there may be cases where several SCDs have been better before the update and others not, or ReFrESH has been run very often over a long period of time.
	In both cases, some SCDs and their associations to referenced sentences may somehow degenerate.
	Then, a user may want to revert some SCDs while keeping others.
	To detect degenerated SCDs, the factors between SCDs and their referenced sentences generated by ReFrESH can be used.
	In addition, this should also be combined with an automated decision approach similar to IFI, but for pruning degenerate parts of SCDs.

	\section*{Acknowledgment}
		The research was partly funded by the Deutsche Forschungsgemeinschaft (DFG, German Research Foundation) under Germany's Excellence Strategy -- EXC 2176 \enquote{Understanding Written Artefacts: Material, Interaction and Transmission in Manuscript Cultures}, project no.\ 390893796.
		The research was conducted within the scope of the Centre for the Study of Manuscript Cultures (CSMC) at Universit\"at Hamburg.
		
		The authors thank the AI Lab Lübeck for providing the hardware used in the evaluation.

	\bibliographystyle{IEEEtran}
	\bibliography{literature.bib}

\end{document}